\documentstyle[prl,aps,floats,psfig]{revtex} \begin{document}
\twocolumn[
\hsize\textwidth\columnwidth\hsize\csname@twocolumnfalse\endcsname

\draft

\title{Phonon-Induced Spin Relaxation of Conduction Electrons in
Aluminum}

\author{Jaroslav Fabian and S. Das Sarma}

\address{Department of Physics, University of Maryland at College
Park, College Park, Maryland 20742-4111}

\maketitle

\begin{abstract}
Spin-flip Eliashberg function $\alpha_S^2F$
and temperature-dependent spin relaxation time $T_1(T)$ are
calculated for aluminum using realistic pseudopotentials.
The spin-flip electron-phonon coupling constant $\lambda_S$ is
found to be $2.5\times 10^{-5}$. The calculations agree
with experiments validating the Elliott-Yafet theory and the
spin-hot-spot picture of spin relaxation for polyvalent metals.
\end{abstract}
]

Spin dynamics of itinerant electrons in metals and semiconductors
is attracting increasing attention. Part of the reason for
this interest is fundamental, arising from improved spin
injection and detection techniques \cite{fabian99b} which now allow
precise measurements of spin transport, relaxation, and coherence
properties. But much of the recent interest is also motivated by the
exciting potential of using electron spin as a building block in
nanoelectronics (dubbed ``spintronics'') where spin dynamics and
transport is projected to be utilized in proposed novel device
applications. The most ambitious such possibility is using electron
spin as a qubit in a quantum computer architecture, but more modest
proposals involving the use of spin injection and transport in new
quantum transistor devices (``spin transistors'') have also been made
\cite{fabian99b}. 

Electron spin already plays a
fundamental, albeit passive, role in giant magnetoresistance-based 
memory devices. The current push for a better
understanding of spin dynamics in electronic materials is, however,
 based on
the hope that the electron spin could be used as an {\it active}
element, where manipulation of spin in a controlled manner will lead
to novel device applications which are not feasible in
conventional microelectronics. This hope arises from two underlying
concepts: the inherently quantum mechanical nature of spin 
(enabling the possibility of truly quantum devices which could not be
envisioned within standard micro- or nanoelectronics),
and, even more importantly, the inherently long relaxation or
coherence time of spin eigenstates in metals and semiconductors 
(indeed, in a typical nonmagnetic metal at room temperature electron 
spins survive for hundreds of picoseconds; by comparison,   
momentum states live no more than femtoseconds). 
This Letter provides the first realistic quantitative calculation
of the temperature dependent spin relaxation time (the so called
$T_1$ relaxation time) in an electronic material, namely, metallic
aluminum. The calculation, for reasons to be explained below, 
is surprisingly subtle and extremely computationally
demanding; it has therefore never been attempted before 
although the basic theory for the phenomenon goes back
more than thirty-five years \cite{elliott54,yafet63}.

The mechanism behind spin relaxation in metals
is believed to be the spin-flip scattering of electrons off
phonons and impurities, as suggested by Elliott \cite{elliott54} and
Yafet \cite{yafet63}.  The periodic, ion-induced spin-orbit interaction 
causes electronic Bloch states to have both spin up and spin down 
amplitudes. The
states can still be polarized by a magnetic field (so we can
call them up and down) but  because of the spin mixing, even a
spin-independent interaction with phonons or impurities (which are assumed
to be nonmagnetic) leads to
a transition from, say,  up to down, degrading any unbalanced spin
population. (Note that the spin-orbit interaction by itself does not
produce spin relaxation--what is needed is spin-orbit coupling to mix
the up and down spins, and a momentum conservation-breaking mechanism such
as impurities or phonons.) Although these arguments seem to be
consistent with experimental findings, there has been to date no
calculation of $T_1$ for a metal based on the Elliott-Yafet theory.

In this Letter we calculate the phonon contribution to $T_1$ for
aluminum providing the first quantitative justification of the theory.
(Impurities in real samples contribute only a temperature
independent background which can be subtracted from the measurement.)
At temperatures $T$ above 100 K, where experimental data are not
available, our calculation is a prediction which should be
useful for designing room-temperature spintronic devices
that use aluminum.
We also calculate the spin-flip
Eliashberg function $\alpha_S^2 F(\Omega)$ which measures the ability
of phonons with frequency $\Omega$ to change electron momenta {\it
and} spins.  This function, which is an analogue 
of the ordinary (spin-conserving) Eliashberg function 
$\alpha^2F(\Omega)$\cite{grimvall81}, is important in
spin-resolved point-contact spectroscopy where phonon-induced spin
flips could be directly observed. (A recent effort
\cite{lang96} to detect phonon-induced spin flips in aluminum  failed
because of the overwhelming spin-flip boundary
scattering in the sample.)

Aluminum belongs to the group of metals whose spin relaxation is
strongly influenced by band-structure anomalies
\cite{fabian98}. Monod and Beuneu \cite{monod79} observed
that while simple estimates based on the Elliott-Yafet theory
work well for monovalent alkali and noble metals,
they severely underestimate $1/T_1$ for polyvalent Al,
Mg, Be, and Pd (the only polyvalent metals measured so far).
Silsbee and Beuneu \cite{silsbee83} pointed out
that in aluminum accidental degeneracies can significantly enhance
$1/T_1$.  We recently \cite{fabian98} developed a general theory
including band structure anomalies like accidental degeneracy,
crossing Brillouin zone boundaries or special symmetry points, 
and rigorously showed
that they all enhance $1/T_1$. This explains the Monod-Beuneu finding
because the anomalies (which we named ``spin hot
spots''\cite{fabian98}) are ubiquitous in polyvalent metals. The
present calculation is consistent with the spin-hot-spot picture.

The formula for the spin relaxation rate, first derived
by Yafet\cite{yafet63}, can be written in the more conventional
electron-phonon terminology \cite{allen75,grimvall81} as
\begin{eqnarray} \label{eq:reltime}
1/T_1(T) =
8\pi T\int_0^{\infty} d\Omega\, \alpha^2_S F(\Omega)
\frac{\partial N(\Omega)}{\partial T},
\end{eqnarray}
where $N(\Omega)=[\exp(\hbar\Omega/k_BT)-1]^{-1}$ and
$\alpha_S^2F(\Omega)$ is the spin-flip Eliashberg
function. Before writing the expression for $\alpha_S^2F$ we
introduce the following notation. Electron states $\Psi$
(normalized to a primitive cell) in the periodic
potential $V$ containing the spin-orbit interaction are labeled
by lattice momentum ${\bf k}$,
band index $n$, and spin polarization $\uparrow$ or $\downarrow$. If
$V$ has inversion symmetry (as in aluminum), states
$\Psi_{{\bf k}n\uparrow}$ and $\Psi_{{\bf k}n\downarrow}$
are degenerate \cite{elliott54}.  The spin
polarization then means that these two states are chosen to satisfy
$(\Psi_{{\bf k}n\uparrow}, \hat{\sigma}_z \Psi_{{\bf
k}n\uparrow})=-(\Psi_{{\bf k}n\downarrow}, \hat{\sigma}_z \Psi_{{\bf
k}n\downarrow}) > 0 $ with the off-diagonal terms
vanishing \cite{yafet63}.  Lattice vibrations
are represented by phonons with momentum ${\bf q}$
and polarization index $\nu$. Phonon frequency is
$\omega_{{\bf q}\nu}$ and polarization vector
${\bf u}_{{\bf q}\nu}$ (we consider a Bravais lattice).
If ${\bf q}={\bf k}-{\bf k'}$ and
\begin{eqnarray} g^{\nu}_{{\bf
k}n\uparrow,{\bf k'}n'\downarrow} \equiv \left | {\bf u}_{{\bf
q}\nu}\cdot\left(\Psi_{{\bf k}n\uparrow},\nabla V \Psi_{{\bf
k'}n'\downarrow}\right)\right |^2, \end{eqnarray} the spin-flip
Eliashberg function is
\begin{eqnarray}\label{eq:elias} \alpha_S^2
F(\Omega)=\frac{g_S}{2M\Omega}\sum_{\nu}\langle\langle
g^{\nu}_{{\bf k}n\uparrow,{\bf
k'}n'\downarrow}\delta(\omega_{{\bf q},\nu}-\Omega)
\rangle_{{\bf k}n}\rangle_{{\bf k'}n'}.
\end{eqnarray}
Here $g_S$ is the number of states per spin and atom
at the Fermi level, $M$ is the ion mass, and
$\langle ...  \rangle_{{\bf k}n}$ denotes the Fermi surface averaging
\cite{note1}.

\begin{figure}
\centerline{\psfig{file=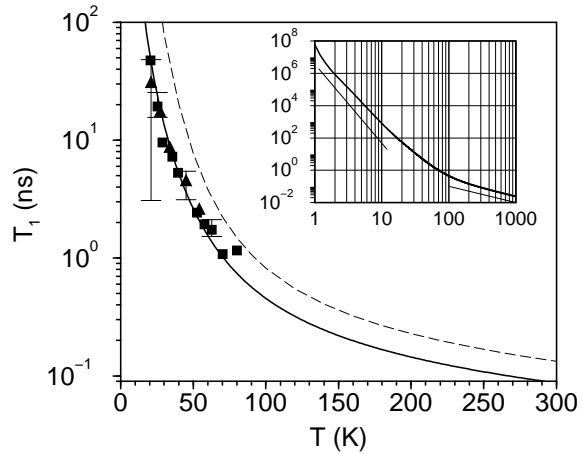,width=0.9\linewidth,angle=0}}
\caption{Calculated spin relaxation time $T_1$ of conduction
electrons in aluminum as a function of temperature $T$  (solid
line). Symbols are experimentally determined \protect\cite{johnson85}
phonon contribution to $T_1$ from measurements by
Johnson and Silsbee \protect\cite{johnson85} (triangles) and Lubzens
and Schultz \protect\cite{lubzens76} (squares).  The dashed line is
an estimate of $T_1$ from Eq. \protect\ref{eq:qualitative}.  The
inset shows $T_1$ over a wider temperature range with thin lines
indicating the predicted low-$T$ ($T_1\sim T^{-5}$) and high-$T$
($T_1\sim T^{-1}$) behavior.
}
\label{fig:1}
\end{figure}

We calculate $\alpha_S^2F$ and $T_1$ for aluminum by the
pseudopotential method \cite{grimvall81}. The spin-independent
part of the electron-ion pseudopotential is represented by the
Ma\v{s}ovi\'{c}-Zekovi\'{c}\cite{masovic78} semi-empirical form
factor which reproduces well the observed band gaps at the symmetry
points of the Brillouin zone.  This is a crucial feature because the
presence of spin hot spots makes $T_1$ sensitive to the band
structure at the Fermi surface \cite{fabian98}.
The spin-orbit part
of the pseudopotential comes from a fit of the first-principles
Bachelet-Hamann-Schl\"{u}ter pseudopotential
\cite{bachelet82} to $\alpha \hat{\bf
L}\cdot\hat{\bf S}{\cal P}_1$, where $\hat {\bf L}$ ($\hat {\bf S}$)
is the orbital (spin) momentum operator and ${\cal P}_l$ is the
operator projecting on the orbital momentum state $l$.  The parameter
$\alpha=2.4\times 10^{-3}$ a.\,u. (1 a.\,u. = 2 Ry)  inside the ion
core of twice the Bohr radius, $r_c=2r_B$.  Outside the core $\alpha$
vanishes. The cutoff for the plane-wave energy is 1 a.\,u.
from the Fermi level \cite{note2}.
For phonons we use the highly successful force-constant model of
Cowley \cite{cowley74} which gives an excellent fit to the
experimental spectrum. Finally, the sums over the Brillouin zone are
done by the tetrahedron method \cite{allen83} with a specially
designed grid of more than 4000 points around the Fermi surface in an
irreducible wedge of the Brillouin zone to accurately obtain
contributions from the spin hot spots.

Figure \ref{fig:1} shows the calculated spin relaxation time $T_1$ as
a function of temperature. The agreement with experiment is evident.
At high temperatures where there are no experimental data, our
calculation predicts $T_1 [{\rm ns}]\approx 24 T^{-1} [K]$.  This
behavior is expected for a phonon-induced relaxation above the Debye
temperature which for aluminum is about 400 K. As Fig. \ref{fig:1}
shows the $T_1\sim T^{-1}$ behavior starts already at 200 K.  At very
low temperatures the theory predicts the asymptotic
temperature dependence $T_1\sim T^{-5}$ (the Yafet law
\cite{yafet63}) purely on dimensional grounds.  Our calculation gives
rather a good fit to $T_1\sim T^{-4.35}$ between 2 and 10 K.  At
lower temperatures our results cease to be reliable because of the
finite size (limited by the computing resources) of the tetrahedron
blocks in the summations over the Brillouin zone.  We anticipate that
the asymptotic Yafet law would be reached at lower temperatures (much lower
than 2 K) since we have verified numerically its origin, namely 
that $g^{\nu}_{{\bf
k'}n\uparrow,{\bf k'}n\downarrow}\sim ({\bf k}-{\bf k'})^4$ at ${\bf
k}\rightarrow {\bf k'}$ \cite{yafet63} (a quadratic dependence would
be expected for spin-conserving matrix elements). 
In Fig.  \ref{fig:1} we also plot an estimate of
$T_1$ based on the simple formula \cite{fabian98}
\begin{eqnarray}
\label{eq:qualitative} T_1\approx \tau/4\langle b^2 \rangle,
\end{eqnarray}
where $\langle b^2 \rangle$ is the Fermi surface
average of the spin-mixing parameter, calculated in \cite{fabian98}
to be $2.5\times 10^{-5}$, and $\tau$ is the momentum relaxation time
obtained  from the Drude formula for the resistivity (resistivity
data taken from Ref.  \cite{seth70}) with an electron thermal mass of
$1.5$ \cite{kittel} of the free electron mass. This estimate of $T_1$
reproduces well the calculated functional temperature dependence
making Eq.  \ref{eq:qualitative} useful as a starting point for
order-of-magnitude estimates.

\begin{figure}
\centerline{\psfig{file=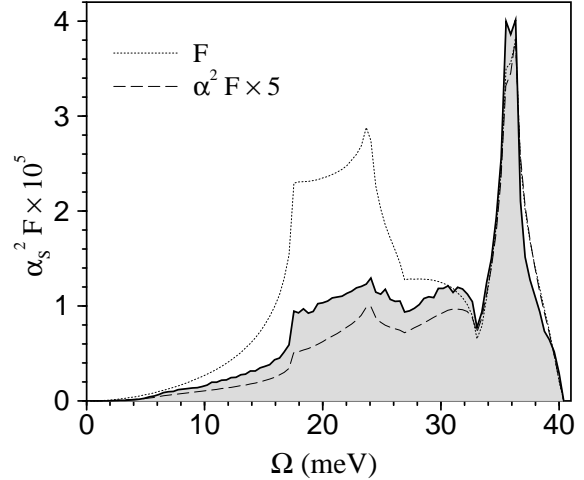,width=\linewidth,angle=0}}
\caption{Spin-flip Eliashberg function $\alpha^2_SF$ for aluminum.
The dotted line shows the phonon density of states $F$ and the dashed
line is the ordinary (spin-conserving) Eliashberg function $\alpha^2
F$. The curves are calculated for the model described in the text.
}
\label{fig:2}
\end{figure}

The calculated spin-flip Eliashberg function $\alpha_S^2F$ for
aluminum is shown in Fig.  \ref{fig:2} along with the phonon density
of states $F$ and the spin-conserving Eliashberg function
$\alpha^2F$. The last agrees very well with previous calculations
\cite{leung76,savrasov96}.  Transverse phonon modes which dominate
the low-frequency spectrum are less effective in scattering
electrons, with or without spin flip, than high-frequency 
longitudinal phonon modes. The behavior of
$\alpha_S^2F$ at small $\Omega$ that gives the Yafet law is predicted
to be $\alpha_S^2F\sim \Omega^4$. We are not able to reproduce this
result, again because of the finite size of the tetrahedron blocks.
This is a well known problem that the asymptotic low-frequency behavior
is hard to reproduce  \cite{leung76,savrasov96}.

From the Eliashberg
function we can calculate the effective electron-phonon coupling
constant \begin{eqnarray} \lambda_{(S)}=2\int_{0}^{\infty}
\frac{d\Omega}{\Omega} \alpha_{(S)}^2F(\Omega).  \end{eqnarray} We
obtain $\lambda\approx 0.4$ and $\lambda_S\approx 2.5\times 10^{-5}$.
The spin-conserving $\lambda$ falls well into the interval of the
``recommended'' values $0.38-0.48$ \cite{grimvall81} obtained by
different methods \cite{leung76,savrasov96,allen69,lam86}.  At high
temperatures the phonon-induced relaxation is determined by
$\lambda_{(S)}$, since in this regime $\hbar/\tau\approx 2\pi\lambda
k_BT$ and $\hbar/T_1\approx 4\pi\lambda_S k_BT$. The momentum to spin
relaxation time ratio $\tau/T_1$ is $2\lambda_S/\lambda\approx 1.24
\times 10^{-4}$. From the above ratio of $\tau/T_1$ we obtain the
``effective'' $\langle b^2 \rangle \approx 3.1 \times 10^{-5}$ in Eq.
\ref{eq:qualitative}, not that different from its calculated value of
$2.0 \times 10^{-5}$ \cite{fabian98}. Thus, our theory is internally
consistent.

We conclude with a remark on the accuracy of our calculation of
$\lambda_S$.  The numerical error is accumulated mostly during the
summations over the Brillouin zone. This error was previously
estimated\cite{savrasov96} to be about 10\%. Another source of uncertainty,
which is much more important here than in the spin-conserving calculations,
comes from the choice of the pseudopotentials. While
the spin-orbit pseudopotential sets the overall scale ($1/T_1\sim
\alpha^2$), the scalar part of the
pseudopotential determines the ``band renormalization'' of $1/T_1$,
that is, the enhancement due to spin hot spots \cite{fabian98}.
Here we can only offer a guess. Considering the
spin-orbit part ``fixed,'' our semi-empirical scalar pseudopotential,
which reproduces the experimental band gaps at symmetry points within
5\%, does not introduce more than another 10\% error\cite{note3},
making $\lambda_S$ determined with 20\% accuracy. As for the
spin-orbit interaction, future experiments done in the regime
where $T_1\sim 1/T$ (that is, above 200 K), will have the opportunity
to set definite constraints on $\alpha$ through a direct comparison
with our theory.

In summary, we have provided the first fully quantitative theory
for the temperature dependent spin relaxation rate in aluminum taking
into account spin-orbit coupling and electron-phonon interaction
within the Elliott-Yafet formalism using realistic pseudopotentials.
Our theoretical results are in excellent agreement with
the measured $T_1(T)$ in aluminum and for $T>$ 100 K, where
experimental results are currently non-existent, our theory provides
specific predictions for comparison with future experiments.

We thank P. B. Allen for helpful discussions. This work
was supported by the U.S. ONR and the U.S. ARO.


\begin{references}
\bibitem{note1} Writing $1/T_1$ as Eq. \ref{eq:reltime}
	    assumes\cite{grimvall81} that $g^{\nu}_{{\bf
	    k}n\uparrow,{\bf k'}n'\downarrow}$ is not
	    altered much by small departures, of order
	    Debye energy, from the Fermi level.
	    This assumption is justified as long as the spin hot
	    spots remain (we checked that this is the case for
	    aluminum).
\bibitem{note2}  In our previous work \cite{fabian98} we used
	     a much larger cutoff of 3 a.\, u. (about 30 plane
	     waves) and $\alpha =2.7 \times 10^{-3}$
	     a.\,u.  inside the ion core. The present value of
	     $\alpha$ is adjusted so that with the lower cutoff
	     of 1 a.\,u.  (about 10 plane waves--this is needed
	     to reduce the computational time) one gets the same
	     average spin-mixing parameter $\langle b^2
	     \rangle$ \cite{fabian98}  (see Eq.
	     \ref{eq:qualitative}).
\bibitem{note3}  An error in symmetry-points band gaps causes the
                same error in $1/T_1$ if spin hot spots are points
                near a Brillouin zone boundary, and twice of the
                error if the spin hot spots are accidental degeneracy
                points \cite{fabian98}.
\end{references}
\end{document}